\documentclass[conference]{IEEEtran}
\usepackage{cite}
\usepackage{graphicx}

\begin{document}


\title{A Cross-Repository Model for Predicting
Popularity in GitHub}

\author{
	
	\IEEEauthorblockN{Neda Hajiakhoond Bidoki\IEEEauthorrefmark{1}, Gita Sukthankar\IEEEauthorrefmark{1}, Heather Keathley\IEEEauthorrefmark{2} and Ivan Garibay\IEEEauthorrefmark{2}}
	\IEEEauthorblockA{
		 \IEEEauthorrefmark{1}Department of Computer Science\\ Email:
		hajiakhoond@cs.ucf.edu, gitars@eecs.ucf.edu\\ \IEEEauthorrefmark{2}Industrial Engineering \& Management Systems\\  Email: heather.keathley@ucf.edu, ivan.garibay@ucf.edu
		}

}
\maketitle

\begin{abstract}
Social coding platforms, such as GitHub, can serve as natural laboratories
for studying the diffusion of innovation through tracking the pattern of code
adoption by programmers. This paper focuses on the problem of predicting the
popularity of software repositories over time; our aim is to forecast the time series
of popularity-related events (code forks and watches). In particular we are interested
in cross-repository patterns—-how do events on one repository affect other
repositories? Our proposed LSTM (Long Short-Term Memory) recurrent neural network
integrates events across multiple active repositories, outperforming a standard
ARIMA (Auto Regressive Integrated Moving Average) time series prediction based
on the single repository. The ability of the LSTM to leverage cross-repository information
gives it a significant edge over standard time series forecasting.
\end{abstract}

\begin{IEEEkeywords}
LSTM, Social Network Analysis, Popularity
\end{IEEEkeywords}
   
\section{Introduction}
As the world becomes more interconnected and project teams are more commonly
geographically dispersed, the role that social networks and social media play in
successful completion of project tasks is quickly becoming accepted in many professional
settings. One example of this is in software development where social networking
services are used to facilitate collaborative development of software code across communities \cite{begel2010social}. GitHub is one of the most commonly used services for asynchronous team-based software development, which provides a space for developers to store source code and interact with formal or informal collaborators
to complete development projects. This platform is relatively unique compared to
other social networks because it brings together professionals who work together
to complete knowledge-based work, which provides an opportunity to investigate
the diffusion of innovation using analytic approaches that leverage the abundance of
data created by activity on GitHub.

Code on GitHub is stored in repositories, and the repository owner and collaborators
make changes to the repository by committing their content. Three event types
in particular are key for tracking public interest in a repository: forking, watching,
and starring. Forks occur when a user clones a repository and becomes its owner.
Sometimes forks are created by the original team of collaborators to manage significant
code changes, but anyone can fork a public repository. Developers can watch
a repository to receive all notifications of changes and star repositories to signal
approval for the project and receive a compressed list of notifications. Forks are
valuable for tracking the spread of innovation, and all three events (fork, watch, and
star) have been used as measures of repository popularity.

In this paper, we demonstrate a repository popularity predictor that can forecast
fork and watch demand for the subset of most active repositories by leveraging cross
repository events. For a given repository, these events can be treated as a sequence
to model the volume of innovation diffusion. For example, Figure 1 shows watch
events corresponding to two different popular repositories on GitHub over a three
year period. Our prediction approach relies on Recurrent Neural Networks (RNNs)
which have been widely used in a variety of sequence learning problems including
unsegmented handwriting generation \cite{graves2013generating} and natural language processing \cite{sutskever2014sequence}.
RNNs can process arbitrary-length sequences of inputs especially when the elements
of the sequence are not independent, i.e., if there exists a hidden relationship among different sequence elements. Here, we employ one of the best performing sequence
learning architectures, Long Short Term Memory (LSTM) \cite{hochreiter1997long}. Our experiments
show that LSTM with cross repository information outperforms an ARIMA
(Auto Regressive Integrated Moving Average) model that forecasts the future events
for a repository using only its own past events. Our evaluation was conducted on a
dataset composed of the public GitHub events and repository profiles from January
2015 through June 2017.

The remainder of this paper is organized as follows. Section \ref{related work} presents the related
work on GitHub and popularity prediction in social media. Section \ref{data} describes
our dataset and our information encoding procedure. The LSTM architecture is introduced
in Section \ref{method}. Results are provided in Section \ref{results} and then we conclude the
paper with a discussion of future work.

\begin{figure*}[h]
\centering\includegraphics[height=6cm,width=15cm]{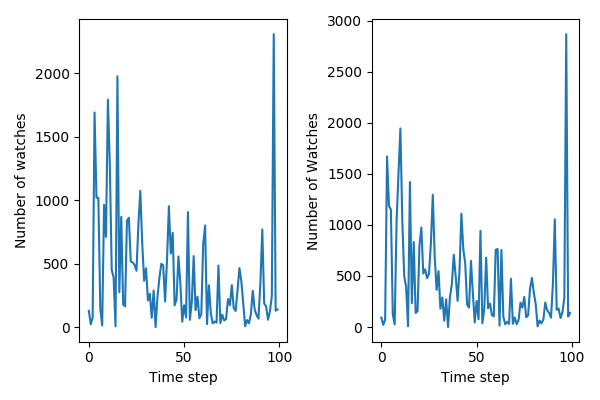}
\caption{Watch event time sequence for two repositories between January 2015 and June 2017. Each
time step represents a ten day period. Note the cross-repository similarity in the volume of events}\label{Fig:repo watch pattern}
\end{figure*}

\section{Related Work}
\label{related work}

Although GitHub is relatively new, there have been many studies conducted on this
social media platform. One locus of interest is understanding social behavior and
teamwork in GitHub communities, using approaches such as regression modeling
to investigate key drivers and behaviors in projects and teams \cite{vasilescu2015gender,Neda}. Ecosystems in
social coding platforms, emerge from commonalities in programming language and
topic, along with code dependencies; it is possible to study their evolution over time
using networks extracted from the GitHub event data \cite{blincoe2015ecosystems,singer2013mutual}.

In addition, there has been research investigating the impact of utilizing social
coding platforms on the software development process \cite{vasilescu2013stackoverflow,gousios2014exploratory,ray2014large,vasilescu2015perceptions}. These studies
highlight the benefits and challenges of completing complex software development
projects in this space. Much of the work in this area utilizes data-driven approaches
that leverage the available data to investigate behaviors such as onboarding, pullrequests,
and documentation evolution\cite{aggarwal2014co,casalnuovo2015developer,soares2018factors}. While many of the studies on
GitHub utilize data-driven techniques to investigate these phenomena, there are also
several examples of survey and interview studies that aim to develop a more nuanced
understanding of these events \cite{dabbish2012social,marlow2013impression,vasilescu2015data}.

Prior work on GitHub popularity prediction demonstrated that the fraction of fork
events a repository has received in the past is an effective heuristic for predicting the
relative distribution of fork events across repositories in the future \cite{gita}. However this
popularity-based model of network evolution was only used to predict the general
structure of the repo-user network rather than future event sequences.

There has also been research on modifying the recurrent neural network architecture
to improve prediction performance. Wu et al. recently introduced a new network
architecture, Deep Temporal Context Networks, for predicting social media
popularity \cite{wu2017sequential}. Rather than using a single time representation, DTCN uses multiple
temporal contexts, combined with a temporal attention mechanism, to improve
performance over a standard LSTM at ranking the popularity of photos on Flickr.
Other types of prediction techniques, such as point process models, have been used to predict tweet popularity, measured by retweeting \cite{zhao2015seismic}. The key contribution of
our paper is illustrating the value of cross-repository information, regardless of the
prediction model employed.

\section{Data Description}
\label{data}
Our GitHub activity dataset consists of 14 event types: CommitComment, Create,
Delete, Fork, Gollum, IssueComment, Issue, Member, Public, Pullrequest, PullrequestReviewComment,
Push, Release, and Watch. These events can be categorized
into three groups: contributions, watches, and forks. This paper only examines
watches and forks since they are the most relevant to repository popularity. The
watch event occurs when a user stars a repository, and the fork event creates a copy
of a repository that the user can modify without changing the original.

Our dataset includes the period from January 2015 to June 2017. First we divide
the time range into ten day periods to be converted into sequences of watch or fork
events. For our study, we selected the 100 repositories with the highest number of
watch and fork events, based on their event profiles. The component event information
from the profile is included as a feature. Comparing components also reveals if
there is an undirected path between repositories. Figure 2 shows our data sequence
structure. $n_t$
is the number of either fork or watch events for each repository. $c_t$
is the
ID of the components to which each repository belongs. The input to the network at
$t$ is $x_t = \{n_t,c_t\}$ and the output $y_t = n_t+1$ is the prediction result.

\begin{figure*}[h]
\centering\includegraphics[width=0.485\textwidth]{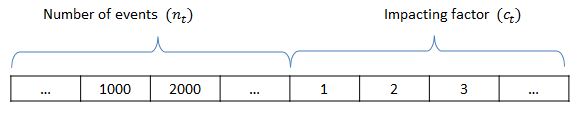}
\caption{Input data structure }
\label{Fig:data_seq}
\end{figure*}

This sequential data is fed to the LSTM neural network in order to learn either
the fork or watch patterns of each repository. After training the model, the prediction
can be made continuously by inputting the new number of associated events in realtime.
For example, number of watches or forks for time-step $t+1$ will be forecast
based on the inputs at time-steps $[1,t-1]$ and the current number of forks or watches
at time-step $t$.

\section{Method}
\label{method}
In GitHub, most users contribute code to multiple repositories and may copy code
from many external repositories. Thus, it is likely that observing the event sequence of one repository may provide information about the user’s activities on other repositories.
Transfer entropy is a measure of influence in social media \cite{ver2012information}; by testing
for transfer entropy (also known as Granger causality) between event sequences, we
observed that fusing information across multiple repository event sequences could
be helpful. To perform this fusion, we needed a model that performs well with multidimensional
time series data in order to simultaneously consider these joint trends;
these considerations guided our choice toward recurrent neural networks (RNNs). In
the context of time series forecasting, RNNs capture and information from the past
inputs and employ them alongside with current input to predict future time steps.
Although, RNNs can store a long sequence of past information theoretically, practically
their memory is limited. Figure 3 illustrates the structure of a general RNN
architecture.

\begin{figure*}[h!]
\centering\includegraphics[width=8.5cm]{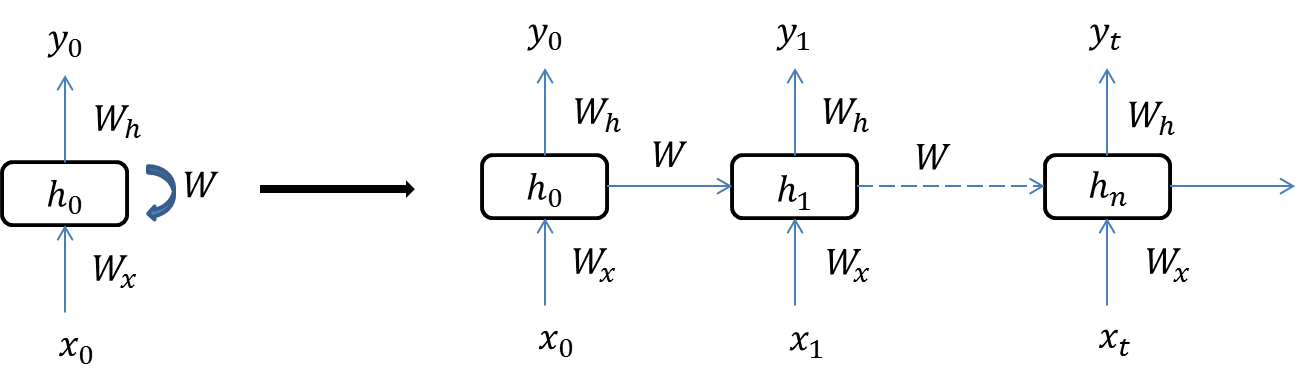}
\caption{Recurrent neural network architecture. The RNN processes input x, stores hidden state h
and outputs y at each time-step t. It then passes the information from one step to the next one. Ws
are the shared weights among different time steps. To train these weights, the network is unfolded
for a finite number of time steps.}\label{Fig:RNN-rolled-unrolled}
\end{figure*}
 \vspace{4pt}
 
 The variables in Figure 3 are as follows:
 
 \begin{itemize}
	\item[-] $x_t$ represents the input data at time-step $t$.
    \item[-] $h_t$ represents the hidden state at time-step $t$, which depends on the previous hidden state as well as the current input.
    \item[-] $y_t$ represents the output at time-step $t$.
    \item[-] $W_{xh}$, $W_{hh}$ and $W_{hy}$ represent shared weights across each unrolled time-step.
\end{itemize}

The central distinguishing feature of RNNs lies in the hidden layer structure.
These layers are in charge of capturing and using the past information from all
previous time-steps. The computations are the same at each time-step, however they
are applied to different inputs $x_t$. Therefore the outputs are different as well. The
shared computation process avoids over-fitting as well as reduces the total number
of parameters. The error is computed based on the difference between the actual
value extracted from the dataset and the concatenation of outputs from all layers. For
this paper, we use the LSTM architecture [10] which is very versatile and has been
shown to perform well on a wide variety of problems  \cite{graves2013generating,sutskever2014sequence}, including prediction
of trends in social media \cite{wu2017sequential}. Our LSTM network model is implemented on top of
Keras.

\begin{table}[ht]
   \centering
\caption{Data set details and hyperparameters}

\begin{tabular}{|l|l|}
\hline
Number of repositories & 100\\
\hline
Time-step length & 10 days\\
\hline
Sequence length & 2/3/4/5/6\\
\hline
Number of features & 100  \\
\hline
Number of hidden layers & 2 \\
\hline
Number of nodes in each hidden layer & 1/2/3 \\
\hline
Loop back & 8 \\
\hline
\end{tabular}
\label{tab:SimulationParameters}
\end{table}

\section{Results}
\label{results}
This section presents an evaluation of our model’s ability to forecast watch and fork
event time series on our dataset, since these events are the best direct measure of
repository popularity. We compare our model to ARIMA (Auto Regressive Integrated
Moving Average). Our data set contains events from January 2015 through
March 2017 for both event types; from this, we sample the 100 repositories with
the highest fork and watch counts during this period. The data from these repositories
was divided into 10 day intervals. 80\% of the training data was used for model
training and the remaining 20\% was reserved for validation. We stop the training
when the validation error does not change for 100 epochs. Figure 4 shows how the
LSTM loop back size was calculated. Table I summarizes the dataset details and
the hyperparameters used in our experiments.

\begin{figure*}[t!]
\centering\includegraphics[width=0.45\textwidth]{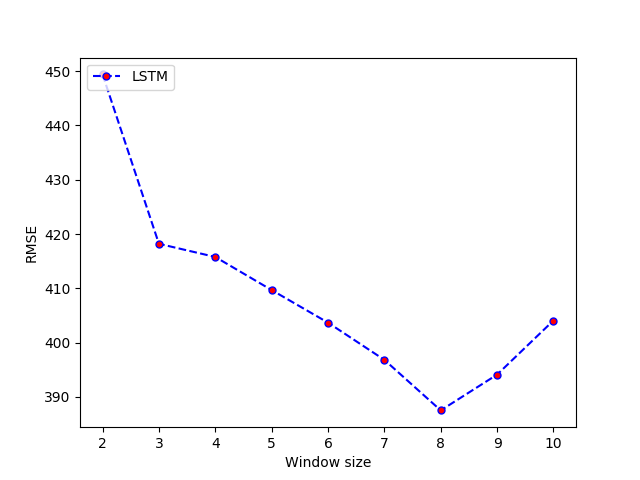}
\caption{LSTM performance vs loop back size. In theory, RNNs can store long sequences but in
practice their memory is limited. Based on these results, we set the loop back size to 8.}
\label{Fig:lopback}
\end{figure*}

\subsection{Benchmarks}
Our benchmark is a standard ARIMA (Auto Regressive Integrated Moving Average)
time series predictor that uses the past time series to forecast the future. To
evaluate our method, the LSTM (with cross-repository information) is compared
to the standard ARIMA model implemented with the Pyramid library, a statistical
Python library that brings R’s auto.arima functionality to Python.

To evaluate our prediction performance, we compared the results of Root Mean
Square Error (RMSE) over time and repositories. $Y_{r,t}$
is the actual number of fork or
watch events for repository $r$ at time-step $t$, and $\widehat{Y}_{r,t}$
is the predicted number of that
type of event. The RMSE for repository r over time $[1,T]$ is:

\begin{equation}
RMSE_r = \sqrt{\frac{1}{T} \sum_{t=1}^{T}\big(Y_{r,t} - \widehat{Y}_{r,t}\big)^2}
\label{equ:RMSE_r}
\end{equation}

To evaluate prediction performance over all repositories, we calculated RMSE of all repositories at time-step t as follows:

\begin{equation}
RMSE_t = \sqrt{\frac{1}{R} \sum_{r=1}^{R}\big(Y_{r,t} - \widehat{Y}_{r,t}\big)^2}
\label{equ:RMSE_t1}
\end{equation}

Here $R$ is the total number of repositories in the dataset. To evaluate the prediction
performance over all repositories, we compare the performance of the LSTM
predictor and ARIMA in terms of $RMSE_r$ and $RMSE_t$ as well as the total average
RMSE as a single value. Table II shows that our proposed method, LSTM with
cross repository information, outperforms the ARIMA time series prediction. The
next section analyzes this result in more detail.

\begin{equation}
Total average of RMSE = \sqrt{\frac{1}{R*T} \sum_{t=1}^{T}\sum_{r=1}^{R}\big(Y_{r,t} - \widehat{Y}_{r,t}\big)^2}
\label{equ:RMSE_t}
\end{equation}

\begin{table}[ht]
   \centering
    \caption{Total Average RMSE for LSTM and ARIMA.
    LSTM Yields Lower Error on the Test
    Data.}
\begin{center}
\begin{tabular}{|l|c|r|}
	\hline
	Model & Average $RMSE$ \\
	\hline
    LSTM & \textbf{312.07}\\
	\hline
    ARIMA & 401.35\\
	\hline
\end{tabular}
\end{center}
\label{table}
\end{table}

\subsection{Analysis}
Figures 5 and 7 present a breakdown of the prediction of watch and fork events
according to the $RMSE_t$ metric, in which the performance of all repositories is averaged
together. LSTM consistently exhibits a lower error over all time steps. The
performance breakdown per repository for the $RMSE_r$ metric is less clear (Figure 6
and 8). The LSTM appears to be better at predicting the global event changes across
all repositories over time; ARIMA is unable to capture this since it lacks the cross
repository features. Figure 9 shows the specific predictions made by ARIMA and
LSTM for each of the repositories; LSTM tends to predict more activity for the
repositories with ARIMA predicting less.
\begin{figure}[h!]
    \centering
    \includegraphics[height=7cm,width=8.5cm]{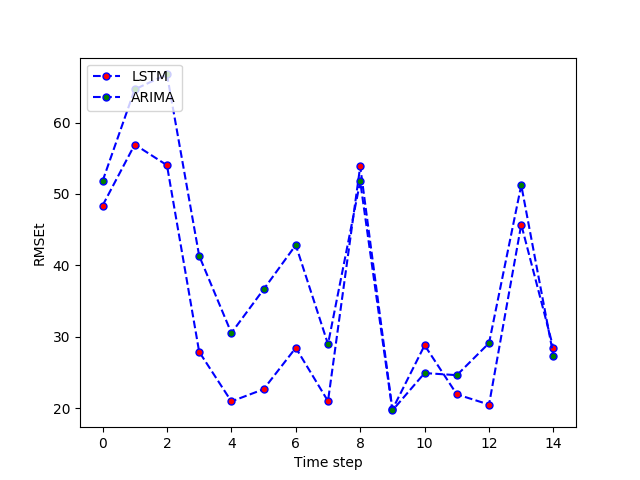}
    \caption{Watch event prediction
performance of two approaches
according to $MSE_t$}
    \label{Fig:watch-MSEt}
\end{figure}

\begin{figure}[h!]
    \centering
    \includegraphics[height=7cm,width=8.5cm]{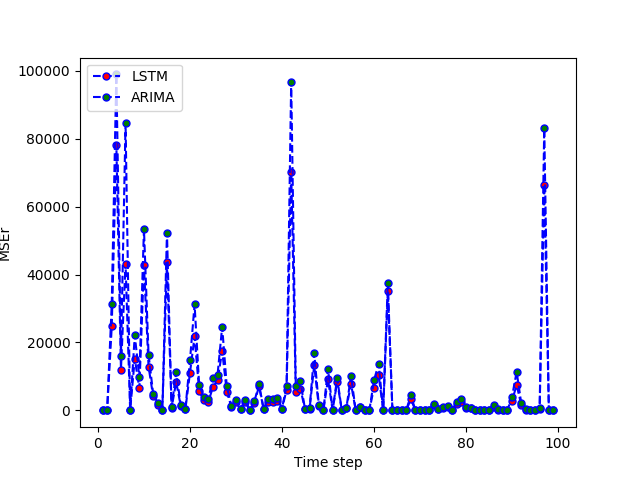}
    \caption{Watch event prediction
performance of two approaches
according to $MSE_r$}
    \label{Fig:fork-sMAPE1}
\end{figure}

\begin{figure}[h!]
    \centering
    \includegraphics[height=7cm,width=8.5cm]{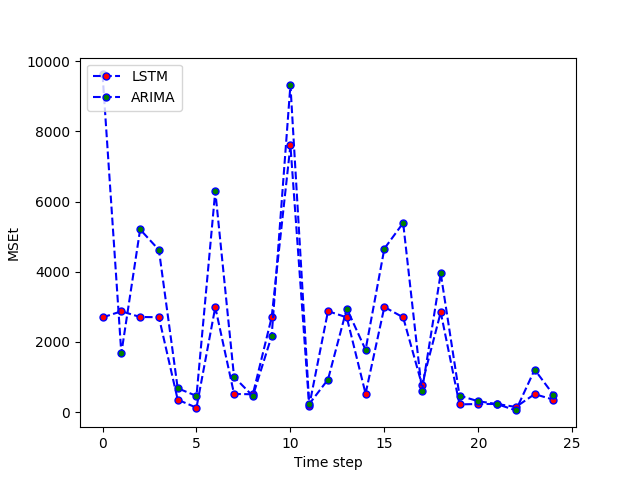}
    \caption{Fork event prediction performance
of two approaches according
to $MSE_t$}
    \label{Fig:watch-sMAPE}
\end{figure}

\begin{figure}[h!]
    \centering
    \includegraphics[height=7cm,width=8.5cm]{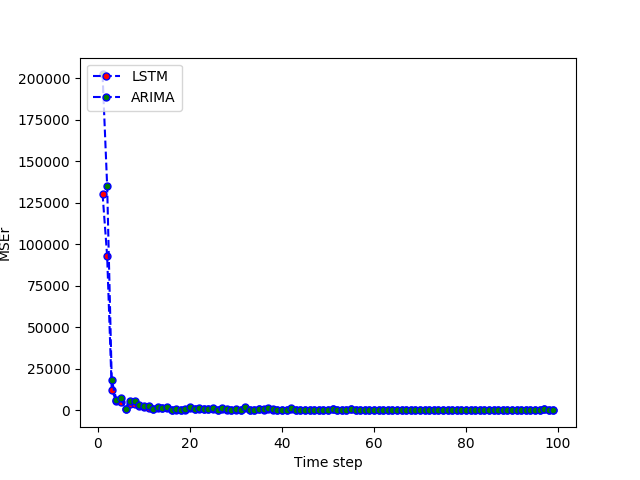}
    \caption{Fork event prediction performance
of two approaches according
to $MSE_r$}
    \label{Fig:fork-sMAPE}
\end{figure}

\begin{figure*}[h]
\centering\includegraphics[height=7cm,width=8.5cm]{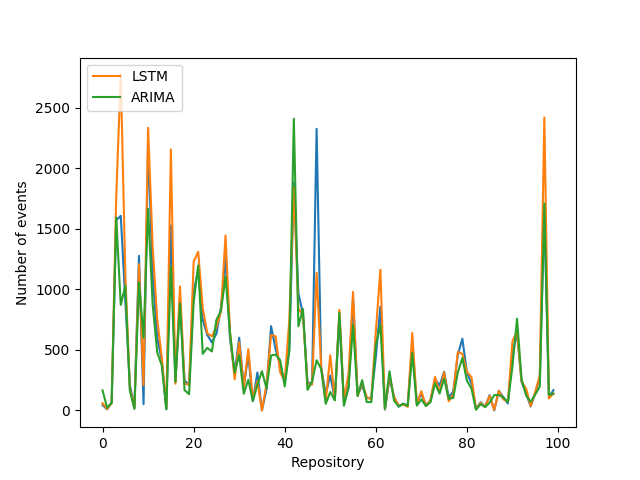}
\caption{Prediction of models across different repositories. LSTM tends to predict more activity for the repositories with ARIMA predicting less.}
\label{Fig:whole_city_smape}
\end{figure*}

\section{Conclusion}
This paper introduces an approach for leveraging cross-repository activity to forecast
trends in GitHub repository popularity. We present results on incorporating
cross-repository features into an LSTM sequence learning model that demonstrate
that it outperforms a standard time series forecast done with ARIMA for predicting
the timing of fork and watch events. The LSTM is clearly better at capturing
general shifts in user activity across GitHub. In future work, we plan to combine
the prediction of multiple event types into a single learned model to come up with
a more cohesive measure of repository popularity. Also we believe that including
past information about specific transfer entropy between repositories could further
improve the prediction performance.

\section*{Acknowledgments}
This research was supported by DARPA program HR001117S0018.

\bibliographystyle{IEEEtran}
 \bibliography{main}

\begin{thebibliography}{10}
\providecommand{\url}[1]{#1}
\csname url@samestyle\endcsname
\providecommand{\newblock}{\relax}
\providecommand{\bibinfo}[2]{#2}
\providecommand{\BIBentrySTDinterwordspacing}{\spaceskip=0pt\relax}
\providecommand{\BIBentryALTinterwordstretchfactor}{4}
\providecommand{\BIBentryALTinterwordspacing}{\spaceskip=\fontdimen2\font plus
\BIBentryALTinterwordstretchfactor\fontdimen3\font minus
  \fontdimen4\font\relax}
\providecommand{\BIBforeignlanguage}[2]{{%
\expandafter\ifx\csname l@#1\endcsname\relax
\typeout{** WARNING: IEEEtran.bst: No hyphenation pattern has been}%
\typeout{** loaded for the language `#1'. Using the pattern for}%
\typeout{** the default language instead.}%
\else
\language=\csname l@#1\endcsname
\fi
#2}}
\providecommand{\BIBdecl}{\relax}
\BIBdecl

\bibitem{begel2010social}
A.~Begel, R.~DeLine, and T.~Zimmermann, ``Social media for software
  engineering,'' in \emph{Proceedings of the FSE/SDP Workshop on Future of
  Software Engineering Research}.\hskip 1em plus 0.5em minus 0.4em\relax ACM,
  2010, pp. 33--38.

\bibitem{graves2013generating}
A.~Graves, ``Generating sequences with recurrent neural networks,'' \emph{arXiv
  preprint arXiv:1308.0850}, 2013.

\bibitem{sutskever2014sequence}
I.~Sutskever, O.~Vinyals, and Q.~V. Le, ``Sequence to sequence learning with
  neural networks,'' in \emph{Proceedings of NIPS}, December 2014, pp.
  3104--3112.

\bibitem{hochreiter1997long}
S.~Hochreiter and J.~Schmidhuber, ``Long short-term memory,'' \emph{Neural
  Computation}, vol.~9, no.~8, pp. 1735--1780, 1997.

\bibitem{vasilescu2015gender}
B.~Vasilescu, D.~Posnett, B.~Ray, M.~G. van~den Brand, A.~Serebrenik,
  P.~Devanbu, and V.~Filkov, ``Gender and tenure diversity in {GitHub} teams,''
  in \emph{Proceedings of the Annual Conference on Human Factors in Computing
  Systems}.\hskip 1em plus 0.5em minus 0.4em\relax ACM, 2015, pp. 3789--3798.

\bibitem{Neda}
N.~Hajiakhoond~Bidoki and G.~Sukthankar, ``A communication protocol for
  man-machine networks,'' in \emph{arXiv: 1808.07975 [cs. MA]}, 2018, pp.
  1513--1522.

\bibitem{blincoe2015ecosystems}
K.~Blincoe, F.~Harrison, and D.~Damian, ``Ecosystems in {GitHub} and a method
  for ecosystem identification using reference coupling,'' in \emph{Proceedings
  of the Working Conference on Mining Software Repositories}.\hskip 1em plus
  0.5em minus 0.4em\relax IEEE Press, 2015, pp. 202--207.

\bibitem{singer2013mutual}
L.~Singer, F.~Figueira~Filho, B.~Cleary, C.~Treude, M.-A. Storey, and
  K.~Schneider, ``Mutual assessment in the social programmer ecosystem: an
  empirical investigation of developer profile aggregators,'' in
  \emph{Proceedings of the ACM Conference on Computer Supported Cooperative
  Work}, 2013, pp. 103--116.

\bibitem{vasilescu2013stackoverflow}
B.~Vasilescu, V.~Filkov, and A.~Serebrenik, ``Stackoverflow and {GitHub}:
  Associations between software development and crowdsourced knowledge,'' in
  \emph{Proceedings of the International Conference on Social Computing}, 2013,
  pp. 188--195.

\bibitem{gousios2014exploratory}
G.~Gousios, M.~Pinzger, and A.~v. Deursen, ``An exploratory study of the
  pull-based software development model,'' in \emph{Proceedings of the
  International Conference on Software Engineering}.\hskip 1em plus 0.5em minus
  0.4em\relax ACM, 2014, pp. 345--355.

\bibitem{ray2014large}
B.~Ray, D.~Posnett, V.~Filkov, and P.~Devanbu, ``A large scale study of
  programming languages and code quality in {GitHub},'' in \emph{Proceedings of
  the ACM SIGSOFT International Symposium on Foundations of Software
  Engineering}, 2014, pp. 155--165.

\bibitem{vasilescu2015perceptions}
B.~Vasilescu, V.~Filkov, and A.~Serebrenik, ``Perceptions of diversity on
  {GitHub}: A user survey,'' in \emph{Proceedings of the Eighth International
  Workshop on Cooperative and Human Aspects of Software Engineering}.\hskip 1em
  plus 0.5em minus 0.4em\relax IEEE Press, 2015, pp. 50--56.

\bibitem{aggarwal2014co}
K.~Aggarwal, A.~Hindle, and E.~Stroulia, ``Co-evolution of project
  documentation and popularity within {GitHub},'' in \emph{Proceedings of the
  Working Conference on Mining Software Repositories}.\hskip 1em plus 0.5em
  minus 0.4em\relax ACM, 2014, pp. 360--363.

\bibitem{casalnuovo2015developer}
C.~Casalnuovo, B.~Vasilescu, P.~Devanbu, and V.~Filkov, ``Developer onboarding
  in {GitHub}: the role of prior social links and language experience,'' in
  \emph{Proceedings of the Joint Meeting on Foundations of Software
  Engineering}.\hskip 1em plus 0.5em minus 0.4em\relax ACM, 2015, pp. 817--828.

\bibitem{soares2018factors}
D.~M. Soares, M.~L. de~Lima~J{\'u}nior, A.~Plastino, and L.~Murta, ``What
  factors influence the reviewer assignment to pull requests?''
  \emph{Information and Software Technology}, vol.~98, pp. 32--43, 2018.

\bibitem{dabbish2012social}
L.~Dabbish, C.~Stuart, J.~Tsay, and J.~Herbsleb, ``Social coding in {GitHub}:
  transparency and collaboration in an open software repository,'' in
  \emph{Proceedings of the ACM Conference on Computer Supported Cooperative
  Work}, 2012, pp. 1277--1286.

\bibitem{marlow2013impression}
J.~Marlow, L.~Dabbish, and J.~Herbsleb, ``Impression formation in online peer
  production: activity traces and personal profiles in {Github},'' in
  \emph{Proceedings of the ACM Conference on Computer Supported Cooperative
  Work}, 2013, pp. 117--128.

\bibitem{vasilescu2015data}
B.~Vasilescu, A.~Serebrenik, and V.~Filkov, ``A data set for social diversity
  studies of {GitHub} teams,'' in \emph{Proceedings of the Working Conference
  on Mining Software Repositories}.\hskip 1em plus 0.5em minus 0.4em\relax IEEE
  Press, 2015, pp. 514--517.

\bibitem{gita}
A.~Al-Rubaye and G.~Sukthankar, ``A popularity-based model of the diffusion of
  innovation on {GitHub},'' in \emph{Proceedings of the Computational Social
  Science Society of the Americas}, Santa Fe, NM, 2018.

\bibitem{wu2017sequential}
B.~Wu, W.-H. Cheng, Y.~Zhang, Q.~Huang, J.~Li, and T.~Mei, ``Sequential
  prediction of social media popularity with deep temporal context networks,''
  in \emph{Proceedings of the International Joint Conference on Artificial
  Intelligence}, July 2017.

\bibitem{zhao2015seismic}
Q.~Zhao, M.~Erdogdu, H.~He, A.~Rajaraman, and J.~Leskovec, ``Seismic, a
  self-exciting point process model for predicting tweet popularity,'' in
  \emph{Proceedings of KDD}, 2015.

\bibitem{ver2012information}
G.~Ver~Steeg and A.~Galstyan, ``Information transfer in social media,'' in
  \emph{Proceedings of the International Conference on World Wide Web}, ser.
  WWW.\hskip 1em plus 0.5em minus 0.4em\relax New York, NY, USA: ACM, 2012, pp.
  509--518.

\end{thebibliography}


@inproceedings{Neda,
  title={A Communication Protocol for Man-Machine Networks},
  author={Hajiakhoond Bidoki, Neda and Sukthankar, Gita},
  booktitle={ arXiv: 1808.07975 [cs. MA]},
  pages={1513--1522},
  year={2018},
  organization={}
}



@inproceedings{ver2012information,
 author = {Ver Steeg, Greg and Galstyan, Aram},
 title = {Information Transfer in Social Media},
 booktitle = {Proceedings of the International Conference on World Wide Web},
 series = {WWW },
 year = {2012},
 pages = {509--518},
 publisher = {ACM},
 address = {New York, NY, USA},
} 

@inproceedings{gita,
   author = "Abduljaleel Al-Rubaye and Gita Sukthankar",
   title = "A Popularity-Based Model of the Diffusion of Innovation on {GitHub}",
   booktitle = "Proceedings of the Computational Social Science Society of the Americas",
   address = "Santa Fe, NM",
   year = "2018"
}


@inproceedings{wu2017sequential,
  title={Sequential Prediction of Social Media Popularity with Deep Temporal Context Networks},
  author={Bo Wu and Wen-Huang Cheng and Yongdong Zhang and Qiushi Huang and Jintao Li and Tao Mei},
  booktitle = "Proceedings of the International Joint Conference on Artificial Intelligence",
  month = "July",
  year = "2017",
}
@inproceedings{zhao2015seismic,
  title="Seismic, a self-exciting point process model for predicting tweet popularity",
  author="Qingyuan Zhao and Murat Erdogdu and Hera He and Anand Rajaraman and Jure Leskovec",
  booktitle = "Proceedings of KDD",
  year = "2015",
}


@article{Van2015,
  author={{B. Van Merriënboer, D. Bahdanau, V. Dumoulin, D. Serdyuk, D. Warde-Farley, J. Chorowski and Y. Bengio}},
  title={Blocks and fuel: Frameworks for deep learning},
  journal={arXiv preprint arXiv:1506.00619},
  year={2015}
}


@online{nyc_taxi,
  author={{NYC Taxi Limousine Commission}},
  title= {Taxi and Limousine Commission (TLC) Trip Record Data},
  url= {http://www.nyc.gov/html/tlc/html/about/trip\_record\_data.shtml}
}

@article{al2017path,
  title={Path planning for mobile DCs in future cities},
  author={Al-Turjman, Fadi and Karakoc, Mehmet and Gunay, Melih},
  journal={Annals of Telecommunications},
  volume={72},
  number={3-4},
  pages={119--129},
  year={2017},
  publisher={Springer}
}

@article{Moreira-Matias2013,
  title={Predicting taxi passenger demand using streaming data},
  author={Luis Moreira-Matias and Joao Gama and Michel Ferreira and Joao Mendes-Moreira and Luis Damas},
  journal={IEEE Transactions on Intelligent Transportation Systems},
  volume={14},
  number={3},
  pages={1393-1402},
  year={2013},
}

@article{feimiao2016,
  title={Taxi Dispatch With Real-Time Sensing Data in Metropolitan Areas: A Receding Horizon Control Approach},
  author={Miao, Fei and Shuo Han and Shan Lin and John A. Stankovic and Desheng Zhang and Sirajum Munir and Hua Huang and Tian He and George J. Pappas},
  journal={IEEE Transactions on Automation Science and Engineering},
  volume={13},
  number={2},
  pages={463-478},
  year={2016},
}


@article{TTD2016,
  title={Theano: A Python framework for fast computation of mathematical expressions},
  author={{The Theano Development}},
  journal={arXiv preprint arXiv:1605.02688},
  year={2016}
}



@inproceedings{kaizhang2016,
  title={A Framework for Passengers Demand Prediction and Recommendation},
  author={Kai Zhang and Zhiyong Feng and Shizhan Chen and Keman Huang and Guiling Wang},
  pages={340--347}, 
  booktitle = "Proc. of IEEE SCC'16",
  month = "June",
  year = "2016",
}

@article{simonyan2014very,
  title={Very deep convolutional networks for large-scale image recognition},
  author={Simonyan, Karen and Zisserman, Andrew},
  journal={arXiv preprint arXiv:1409.1556},
  year={2014}
}

@inproceedings{sutskever2014sequence,
  title={Sequence to sequence learning with neural networks},
  author={Sutskever, Ilya and Vinyals, Oriol and Le, Quoc V},
  booktitle={Proceedings of NIPS},
  pages={3104--3112},
  month = "December",
  year={2014}
}

@book{bishop1994mixture,
  title={Mixture density networks},
  author={Bishop, Christopher M},
  year={1994},
  publisher={Aston University}
}

@inproceedings{zhao2016,
   author = {Kai Zhao and Denis Khryashchev and Juliana Freire and Cláudio Silva and Huy Vo},
   title = "Predicting taxi demand at high spatial resolution: Approaching the limit of predictability",
   booktitle = "Proc. of IEEE BigData'16",
   month = "December",
   year = "2016",
   pages="833--842",
}



@inproceedings{Donahue2015,
   author = {Jeffrey Donahue and Lisa Anne Hendricks and Sergio Guadarrama and Marcus Rohrbach and Subhashini Venugopalan and Kate Saenko and Trevor Darrell},
   title = "Long-term recurrent convolutional networks for visual recognition and description",
   booktitle = "Proc. of IEEE cvpr'15",
   month = "June",
   year = "2015",
   pages="2625--2634",
}


@inproceedings{Alex2013speech,
   author = {Alex Graves and Abdel-rahman Mohamed and Geoffrey Hinton},
   title = "Speech recognition with deep recurrent neural networks",
   booktitle = "Proc. of IEEE icassp'13",
   month = "May",
   year = "2013",
   pages="6645--6649",
}

@book{mclachlan1988mixture,
    author = {McLachlan, Geoffrey J and Basford, Kaye E},
    publisher = {New York: Marcel Dekker},
    title = {Mixture models: Inference and applications to clustering},
    volume={84},
    year = {1988}
}


@article{hochreiter1997long,
  title={Long short-term memory},
  author={Hochreiter, Sepp and Schmidhuber, J{\"u}rgen},
  journal={Neural Computation},
  volume={9},
  number={8},
  pages={1735--1780},
  year={1997},
  publisher={MIT Press}
}

@article{Gonzales14,
author={EJ. Gonzales and J. Yang and EF. Morgul and K. Ozbay},
title={Modeling taxi demand with GPS data from taxis and transit},
journal={MNTRC Report},
volume={12-16},
year={2014},
}


@article{graves2013generating,
  title={Generating sequences with recurrent neural networks},
  author={Alex Graves},
  journal={arXiv preprint arXiv:1308.0850},
  year={2013},
}

@article{Rahmatizadeh2016,
  title={Learning real manipulation tasks from virtual demonstrations using LSTM},
  author={Rouhollah Rahmatizadeh and Pooya Abolghasemi and Aman Behal and Ladislau B{\"o}l{\"o}ni},
  journal={arXiv preprint arXiv:1603.03833},
  year={2016},
}



@article{Karpathy2015,
  title={Visualizing and understanding recurrent networks},
  author={Andrej Karpathy and Justin Johnson and Fei-Fei Li},
  journal={arXiv preprint arXiv:1506.02078},
  year={2015},
}

@ONLINE{geohash,
  author = {Gustavo Niemeyer},
  title = {Tips \& Tricks about geohash},
  year = {2008}, 
  url = {http://geohash.org/site/tips.html}
}


@article{zhang2016trans,
  title={Taxi-Passenger-Demand Modeling Based on Big Data from a Roving Sensor Network},
  author={Desheng Zhang and Tian He and Shan Lin and Sirajum Munir and John A. Stankovic},
  journal={IEEE Transactions on Big Data},
  volume={PP},
  number={99},
  pages={1-1},
  year={2016},
}

@article{cosic2013interpreting,
  title={Interpreting development of unmanned aerial vehicles using systems thinking},
  author={{\'C}osi{\'c}, Jelena and {\'C}urkovi{\'c}, Petar and Kasa{\'c}, Josip and Stepani{\'c}, Josip},
  journal={Interdisciplinary Description of Complex Systems},
  volume={11},
  number={1},
  pages={143--152},
  year={2013},
  publisher={Hrvatsko interdisciplinarno dru{\v{s}}tvo}
}

@inproceedings{Rahmatizadeh-2015-GLOBECOM,
   author = "R. Rahmatizadeh and S. Khan and A.P. Jayasumana and D. Turgut and L. B{\"o}l{\"o}ni",
   title = "Circular update directional virtual coordinate routing protocol in sensor networks",
   booktitle = "Proc. of IEEE GLOBECOM'15",
   month = "December",
   year = "2015",
}


@inproceedings{Solmaz-2014-GLOBECOM,
   author = "G. Solmaz and K. Akkaya and D. Turgut",
   title = "Communication-constrained p-center Problem for Event Coverage in Theme Parks",
   booktitle = "Proc. of IEEE GLOBECOM'14",
   month = "December",
   year = "2014",
   pages="486--491",
}


@inproceedings{akbas2010fapebook,
author={Akba\c{s}, Mustafa  {\.{I}}lhan and Brust, Matthias R. and Ribeiro, Carlos H.C. and Turgut, Damla},
title={{fAPEbook}-Animal Social Life Monitoring with Wireless Sensor and Actor Networks},
booktitle={Proc. of IEEE GLOBECOM'10},
pages={1--5},
month={December},
year={2010},
}


@inproceedings{brust2015networked,
author={Brust, Matthias R. and Strimbu, Bogdan M.},
title={A networked swarm model for {UAV} deployment in the assessment of forest environments},
booktitle={Proc. of IEEE ISSNIP'15},
pages={1--6},
month={April},
year={2015},
}

@article{Juang02,
author={Juang, Philo and Oki, Hidekazu and Wang, Yong and Martonosi, Margaret and Peh, Li Shiuan and Rubenstein, Daniel},
title={Energy-efficient computing for wildlife tracking: Design tradeoffs and early experiences with {ZebraNet}},
journal={SIGARCH Comput. Archit. News},
volume={30},
number={5},
pages={96--107},
month={December},
year={2002},
}

@inproceedings{Anthony12,
author={Anthony, David and Bennett, William P. and Vuran, Mehmet C. and Dwyer, Matthew B. and Elbaum, Sebastian and Lacy, Anne and Engels, Mike and Wehtje, Walter},
title={Sensing through the continent: towards monitoring migratory birds using cellular sensor networks},
booktitle={Proc. of ACM IPSN'12},
pages={329--340},
month={April},
year={2012},
}

@article{Aslan12,
author={Aslan, Yunus Emre and Korpeoglu, Ibrahim and Ulusoy, {\"O}zg{\"u}r},
title={A framework for use of wireless sensor networks in forest fire detection and monitoring},
journal={Computers, Environment and Urban Systems},
volume={36},
number={6},
pages={614--625},
month={November},
year={2012},
}

@article{Diaz11,
author={D{\'i}az, Soledad Escolar and P{\'e}rez, Jes{\'u}s Carretero and Mateos, Alejandro Calder{\'o}n and Marinescu, Maria-Cristina and Guerra, Borja Bergua},
title={A novel methodology for the monitoring of the agricultural production process based on wireless sensor networks},
journal={Computers and Electronics in Agriculture},
volume={76},
number={2},
pages={252--265},
month={May},
year={2011}
}

@inproceedings{mainwaring02,
author={Mainwaring, Alan and Culler, David and Polastre, Joseph and Szewczyk, Robert and Anderson, John},
title={Wireless sensor networks for habitat monitoring},
booktitle={Proc. of ACM WSNA'02 Workshop},
pages={88--97},
month={September},
year={2002},
}

@inproceedings{turgut2013ive,
author={{D. Turgut and L. B{\"o}l{\"o}ni}},
title={{IVE}: improving the value of information in energy-constrained intruder tracking sensor networks},
booktitle={Proc. of IEEE ICC'13},
pages={6360--6364},
month={June},
year={2013},
}

@inproceedings{basagnimaximizing,
author={Basagni, Stefano and B{\"o}l{\"o}ni, Ladislau and Gjanci, Petrika and Petrioli, Chiara and Phillips, Cynthia A. and Turgut, Damla},
title={Maximizing the Value of Sensed Information in Underwater Wireless Sensor Networks via an Autonomous Underwater Vehicle},
booktitle={Proc. of IEEE INFOCOM'14},
pages={988--996},
month={April},
year={2014},
}

@inproceedings{khan2014greedy,
author={Khan, Fahad Ahmad and Khan, Saad Ahmad and Turgut, Damla and B{\"o}l{\"o}ni, Ladislau},
title={Greedy path planning for maximizing value of information in underwater sensor networks},
booktitle={Proc. of IEEE LCN'14 Workshops},
pages={610--615},
month={September},
year={2014},
}

@inproceedings{boloni2013scheduling,
author={B{\"o}l{\"o}ni, Ladislau and Turgut, Damla and Basagni, Stefano and Petrioli, Chiara},
title={Scheduling data transmissions of underwater sensor nodes for maximizing value of information},
booktitle={Proc. of IEEE Globecom'13},
pages={438--443},
month={December},
year={2013},
}

@inproceedings{turgut2012pragmatic,
author={Turgut, Damla and B{\"o}l{\"o}ni, Ladislau},
title={A pragmatic value-of-information approach for intruder tracking sensor networks},
booktitle={Proc. of IEEE ICC'12},
pages={4931--4936},
month={June},
year={2012},
}

@article{watkins1992q,
author={Watkins, Christopher J.C.H. and Dayan, Peter},
title={Q-learning},
journal={Machine learning},
volume={8},
number={3-4},
pages={279--292},
month={May},
year={1992},
}

@article{yeow2007energy,
author={Yeow, Wai-Leong and Tham, Chen-Khong and Wong, Wai-Choong},
title={Energy efficient multiple target tracking in wireless sensor networks},
journal={IEEE Transactions on Vehicular Technology},
volume={56},
number={2},
pages={918--928},
month={March},
year={2007},
}

@article{hodgson2013unmanned,
author={Hodgson, Amanda and Kelly, Natalie and Peel, David},
title={Unmanned aerial vehicles {(UAVs)} for surveying marine fauna: a dugong case study},
journal={PloS one},
volume={8},
number={11},
pages={e79556},
month={November},
year={2013},
}


@article{chamoso2014uavs,
author={Chamoso, Pablo and Raveane, William and Parra, Victor and Gonz{\'a}lez, Ang{\'e}lica},
title={{UAVs} Applied to the Counting and Monitoring of Animals},
journal={Ambient Intelligence - Software and Applications},
volume={291},
number={},
pages={71-80},
month={May},
year={2014},
}


@article{chen2004dynamic,
author={Chen, Wei-Peng and Hou, Jennifer C. and Sha, Lui},
title={Dynamic clustering for acoustic target tracking in wireless sensor networks},
journal={IEEE Transactions on Mobile Computing},
volume={3},
number={3},
pages={258--271},
month={July},
year={2004}
}

@inproceedings{rahmatizadeh2014routing,
author={Rahmatizadeh, Rouhollah and Khan, Saad Ahmad and Jayasumana, Anura P. and Turgut, Damla and B{\"o}l{\"o}ni, Ladislau},
title={Routing towards a mobile sink using virtual coordinates in a wireless sensor network},
booktitle={Proc. of IEEE ICC'14},
pages={12--17},
month={June},
year={2014}
}


@article{luo2006mobiroute,
author={Luo, Jun and Panchard, Jacques and Pi{\'o}rkowski, Micha{\l} and Grossglauser, Matthias and Hubaux, Jean-Pierre},
title={MobiRoute: Routing Towards a Mobile Sink for Improving Lifetime in Sensor Networks},
journal={Distributed Computing in Sensor Systems},
volume={4026},
number={},
pages={480--497},
month={June},
year={2006},
}



@inproceedings{akbas2012actor,
author={Akba\c{s}, Mustafa {\.{I}}lhan and Solmaz, G{\'u}rkan and Turgut, Damla},
title={Actor positioning based on molecular geometry in aerial sensor networks},
booktitle={Proc. of IEEE ICC'12},
pages={508--512},
month={June},
year={2012}
}


@inproceedings{basagni2014maximizing,
author={Basagni, Stefano and B{\"o}l{\"o}ni, Ladislau and Gjanci, Petrika and Petrioli, Chiara and Phillips, Cynthia A. and Turgut, Damla},
title={Maximizing the value of sensed information in underwater wireless sensor networks via an autonomous underwater vehicle},
booktitle={Proc. of IEEE INFOCOM'14},
pages={988--996},
month={April},
year={2014}
}

@article{Solmaz-2014-WINET,
author={G{\'u}rkan Solmaz and Damla Turgut},
title={Optimizing Event Coverage in Theme Parks},
journal={Wireless Networks (WINET) Journal},
volume={20},
number={6},
pages={1445-1459},
month={August},
year={2014}
}

@inproceedings{Solmaz-2013-ICC,
author={{G. Solmaz and D. Turgut}},
title={Event Coverage in Theme Parks Using Wireless Sensor Networks with Mobile Sinks},
booktitle={Proc. of IEEE ICC'13},
pages={1522--1526},
month={June},
year={2013}
}

@article{salarian2014energy,
author={Salarian, Hamidreza and Chin, Kwan-Wu and Naghdy, Fazel},
title={An energy-efficient mobile-sink path selection strategy for wireless sensor networks},
journal={IEEE Transactions on Vehicular Technology},
volume={63},
number={5},
pages={2407--2419},
month={June},
year={2014},
}

@inproceedings{Akbas1-2011-LCN,
  author = "Mustafa  {\.{I}}lhan Akba\c{s} and D. Turgut",
  title = "{APAWSAN:} Actor Positioning for Aerial Wireless Sensor and Actor Networks",
  booktitle = "Proc. of IEEE LCN'11",
  year = "2011",
  month = "October",
  pages = "567-574",
}

@inproceedings{Akbas2-2011-LCN,
  author = "Mustafa  {\.{I}}lhan Akba\c{s} and M.R. Brust and C.H.C. Riberio and D. Turgut",
  title = "Deployment and Mobility for Animal Social Life Monitoring Based on
  Preferential Attachment",
  booktitle = "Proc. of IEEE LCN'11",
  year = "2011",
  month = "October",
  pages = "488-495",
}

@misc{chollet2017keras,
  title={Keras Documentation},
  author={Chollet, Fran{\c{c}}ois},
  year={2016}
}

@article{steinberg2015human,
  title={Human performance in a realistic instrument-control task during short-term microgravity},
  author={Steinberg, Fabian and Kalicinski, Michael and Dalecki, Marc and Bock, Otmar},
  journal={PloS one},
  volume={10},
  number={6},
  pages={e0128992},
  year={2015},
  publisher={Public Library of Science}
}
@article{newman2006modularity,
  title={Modularity and community structure in networks},
  author={Newman, Mark EJ},
  journal={Proceedings of the national academy of sciences},
  volume={103},
  number={23},
  pages={8577--8582},
  year={2006},
  publisher={National Acad Sciences}
}
@article{newman2004finding,
  title={Finding and evaluating community structure in networks},
  author={Newman, Mark EJ and Girvan, Michelle},
  journal={Physical review E},
  volume={69},
  number={2},
  pages={026113},
  year={2004},
  publisher={APS}
}
@inproceedings{rosen2004author,
  title={The author-topic model for authors and documents},
  author={Rosen-Zvi, Michal and Griffiths, Thomas and Steyvers, Mark and Smyth, Padhraic},
  booktitle={Proceedings of the 20th conference on Uncertainty in artificial intelligence},
  pages={487--494},
  year={2004},
  organization={AUAI Press}
}
@article{blondel2008fast,
  title={Fast unfolding of communities in large networks},
  author={Blondel, Vincent D and Guillaume, Jean-Loup and Lambiotte, Renaud and Lefebvre, Etienne},
  journal={Journal of statistical mechanics: theory and experiment},
  volume={2008},
  number={10},
  pages={P10008},
  year={2008},
  publisher={IOP Publishing}
}
@article{yin2012latent,
  title={Latent community topic analysis: Integration of community discovery with topic modeling},
  author={Yin, Zhijun and Cao, Liangliang and Gu, Quanquan and Han, Jiawei},
  journal={ACM Transactions on Intelligent Systems and Technology (TIST)},
  volume={3},
  number={4},
  pages={63},
  year={2012},
  publisher={ACM}
}
@inproceedings{berlingerio2011finding,
  title={Finding and characterizing communities in multidimensional networks},
  author={Berlingerio, Michele and Coscia, Michele and Giannotti, Fosca},
  booktitle={Advances in Social Networks Analysis and Mining (ASONAM), 2011 International Conference on},
  pages={490--494},
  year={2011},
  organization={IEEE}
}
@inproceedings{sun2009ranking,
  title={Ranking-based clustering of heterogeneous information networks with star network schema},
  author={Sun, Yizhou and Yu, Yintao and Han, Jiawei},
  booktitle={Proceedings of the 15th ACM SIGKDD international conference on Knowledge discovery and data mining},
  pages={797--806},
  year={2009},
  organization={ACM}
}
@article{ito2018community,
  title={Community Detection and Correlated Attribute Cluster Analysis on Multi-Attributed Graphs},
  author={Ito, Hiroyoshi and Komamizu, Takahiro and Amagasa, Toshiyuki and Kitagawa, Hiroyuki},
  year={2018}
}
@inproceedings{blincoe2015ecosystems,
  title={Ecosystems in {GitHub} and a method for ecosystem identification using reference coupling},
  author={Blincoe, Kelly and Harrison, Francis and Damian, Daniela},
  booktitle={Proceedings of the Working Conference on Mining Software Repositories},
  pages={202--207},
  year={2015},
  organization={IEEE Press}
}
@inproceedings{begel2010social,
  title={Social media for software engineering},
  author={Begel, Andrew and DeLine, Robert and Zimmermann, Thomas},
  booktitle={Proceedings of the FSE/SDP Workshop on Future of Software Engineering Research},
  pages={33--38},
  year={2010},
  organization={ACM}
}
@inproceedings{storey2010impact,
  title={The impact of social media on software engineering practices and tools},
  author={Storey, Margaret-Anne and Treude, Christoph and van Deursen, Arie and Cheng, Li-Te},
  booktitle={Proceedings of the FSE/SDP workshop on Future of software engineering research},
  pages={359--364},
  year={2010},
  organization={ACM}
}
@inproceedings{zhang1996birch,
  title={BIRCH: an efficient data clustering method for very large databases},
  author={Zhang, Tian and Ramakrishnan, Raghu and Livny, Miron},
  booktitle={ACM Sigmod Record},
  volume={25},
  number={2},
  pages={103--114},
  year={1996},
  organization={ACM}
}
@inproceedings{steinbach2000comparison,
  title={A comparison of document clustering techniques},
  author={Steinbach, Michael and Karypis, George and Kumar, Vipin and others},
  booktitle={KDD workshop on text mining},
  volume={400},
  number={1},
  pages={525--526},
  year={2000},
  organization={Boston}
}
@inproceedings{vasilescu2015gender,
  title={Gender and tenure diversity in {GitHub} teams},
  author={Vasilescu, Bogdan and Posnett, Daryl and Ray, Baishakhi and van den Brand, Mark GJ and Serebrenik, Alexander and Devanbu, Premkumar and Filkov, Vladimir},
  booktitle={Proceedings of the Annual Conference on Human Factors in Computing Systems},
  pages={3789--3798},
  year={2015},
  organization={ACM}
}
@inproceedings{calefato2017preliminary,
  title={A preliminary analysis on the effects of propensity to trust in distributed software development},
  author={Calefato, Fabio and Lanubile, Filippo and Novielli, Nicole},
  booktitle={Global Software Engineering (ICGSE), 2017 IEEE 12th International Conference on},
  pages={56--60},
  year={2017},
  organization={IEEE}
}
@inproceedings{vasilescu2013stackoverflow,
  title={Stackoverflow and {GitHub}: Associations between software development and crowdsourced knowledge},
  author={Vasilescu, Bogdan and Filkov, Vladimir and Serebrenik, Alexander},
  booktitle={Proceedings of the International Conference on Social Computing},
  pages={188--195},
  year={2013},
}
@inproceedings{gousios2014exploratory,
  title={An exploratory study of the pull-based software development model},
  author={Gousios, Georgios and Pinzger, Martin and Deursen, Arie van},
  booktitle={Proceedings of the International Conference on Software Engineering},
  pages={345--355},
  year={2014},
  organization={ACM}
}
@article{soares2018factors,
  title={What factors influence the reviewer assignment to pull requests?},
  author={Soares, Daric{\'e}lio M and de Lima J{\'u}nior, Manoel L and Plastino, Alexandre and Murta, Leonardo},
  journal={Information and Software Technology},
  volume={98},
  pages={32--43},
  year={2018},
  publisher={Elsevier}
}

@inproceedings{singer2013mutual,
  title={Mutual assessment in the social programmer ecosystem: an empirical investigation of developer profile aggregators},
  author={Singer, Leif and Figueira Filho, Fernando and Cleary, Brendan and Treude, Christoph and Storey, Margaret-Anne and Schneider, Kurt},
  booktitle={Proceedings of the ACM Conference on Computer Supported Cooperative Work},
  pages={103--116},
  year={2013},
}
@inproceedings{vasilescu2015perceptions,
  title={Perceptions of diversity on {GitHub}: A user survey},
  author={Vasilescu, Bogdan and Filkov, Vladimir and Serebrenik, Alexander},
  booktitle={Proceedings of the Eighth International Workshop on Cooperative and Human Aspects of Software Engineering},
  pages={50--56},
  year={2015},
  organization={IEEE Press}
}
@inproceedings{marlow2013impression,
  title={Impression formation in online peer production: activity traces and personal profiles in {Github}},
  author={Marlow, Jennifer and Dabbish, Laura and Herbsleb, Jim},
  booktitle={Proceedings of the ACM Conference on Computer Supported Cooperative Work},
  pages={117--128},
  year={2013},
}
@inproceedings{dabbish2012social,
  title={Social coding in {GitHub}: transparency and collaboration in an open software repository},
  author={Dabbish, Laura and Stuart, Colleen and Tsay, Jason and Herbsleb, Jim},
  booktitle={Proceedings of the ACM Conference on Computer Supported Cooperative Work},
  pages={1277--1286},
  year={2012},
}
@inproceedings{casalnuovo2015developer,
  title={Developer onboarding in {GitHub}: the role of prior social links and language experience},
  author={Casalnuovo, Casey and Vasilescu, Bogdan and Devanbu, Premkumar and Filkov, Vladimir},
  booktitle={Proceedings of the Joint Meeting on Foundations of Software Engineering},
  pages={817--828},
  year={2015},
  organization={ACM}
}
@inproceedings{aggarwal2014co,
  title={Co-evolution of project documentation and popularity within {GitHub}},
  author={Aggarwal, Karan and Hindle, Abram and Stroulia, Eleni},
  booktitle={Proceedings of the Working Conference on Mining Software Repositories},
  pages={360--363},
  year={2014},
  organization={ACM}
}
@inproceedings{vasilescu2015data,
  title={A data set for social diversity studies of {GitHub} teams},
  author={Vasilescu, Bogdan and Serebrenik, Alexander and Filkov, Vladimir},
  booktitle={Proceedings of the Working Conference on Mining Software Repositories},
  pages={514--517},
  year={2015},
  organization={IEEE Press}
}
@inproceedings{ray2014large,
  title={A large scale study of programming languages and code quality in {GitHub}},
  author={Ray, Baishakhi and Posnett, Daryl and Filkov, Vladimir and Devanbu, Premkumar},
  booktitle={Proceedings of the ACM SIGSOFT International Symposium on Foundations of Software Engineering},
  pages={155--165},
  year={2014},
}
@inproceedings{sculley2010web,
  title={Web-scale k-means clustering},
  author={Sculley, David},
  booktitle={Proceedings of the 19th international conference on World wide web},
  pages={1177--1178},
  year={2010},
  organization={ACM}
}

@misc{octiverse,
  author = {GitHub.com},
  title  = {{The State of the Octoverse} 2017},
  url    = {https://octoverse.github.com/},
}
@inproceedings{Demara-ASEE2018,
  title={{Automated Formation of Peer Learning Cohorts using Computer-Based Assessment Data: A Double-Blind Study within a Software Engineering Course}},
  author={R. F. Demara and D. Turgut and E. Nassiff and S. Bacanli and N. H. Bidoki and J. Xu},
  booktitle={2018 ASEE Annual Conference & Exposition Conference},
  year={2018},
  month={June},
  note = {17 pages},
}
\end{document}